\documentclass[pdflatex,sn-nature]{sn-jnl}

\usepackage{graphicx}%
\usepackage{amsmath,amssymb,amsfonts}%
\usepackage{booktabs}%
\usepackage{xcolor}%
\usepackage{textcomp}%
\usepackage{xspace}%
\usepackage{pgfplots}%
\usepgfplotslibrary{groupplots}%
\usepackage[utf8]{inputenc}
\usepackage{textgreek}
\DeclareUnicodeCharacter{03CC}{\'{o}}
\pgfplotsset{compat=1.17}%

\newcommand{\KTO}{KTaO$_3$\xspace}

\newcommand{\CZOKTO}{CaZrO$_3$/KTaO$_3$\xspace}

\newcommand{\DevA}{Device~S1\xspace}
\newcommand{\DevB}{Device~S2\xspace}
\newcommand{\DevC}{Device~D1\xspace}
\newcommand{\DevD}{Device~D2\xspace}

\begin{document}

\title[KTaO$_3$-based Double Quantum Dots in the single-electron limit]{KTaO$_3$-based Double Quantum Dots in the single-electron limit}

\author[1]{\fnm{Sayanwita} \sur{Biswas}}
\author[2]{\fnm{Jiangfeng} \sur{Yang}}
\author[2]{\fnm{Prithwijit} \sur{Mandal}}
\author[1]{\fnm{Miles} \sur{Klingman}}
\author[1]{\fnm{Sakshi} \sur{Takale}}
\author[1]{\fnm{Ranjani} \sur{Ramachandran}}
\author[1]{\fnm{Aria} \sur{Hajikhani}}
\author[1]{\fnm{Patrick} \sur{Irvin}}
\author[2]{\fnm{Chang-Beom} \sur{Eom}}
\author*[1]{\fnm{Jeremy} \sur{Levy}}\email{jlevy@pitt.edu}

\affil[1]{\orgdiv{Department of Physics and Astronomy}, \orgname{University of Pittsburgh}, \orgaddress{\city{Pittsburgh}, \state{PA}, \country{USA}}}

\affil[2]{\orgdiv{Department of Materials Science and Engineering}, \orgname{University of Wisconsin-Madison}, \orgaddress{\city{Madison}, \state{WI}, \country{USA}}}

\abstract{Reaching the single-electron limit was a turning point for spin-based quantum computation as it moved semiconductor quantum dots from mesoscopic physics to devices in which individual electron spins carry quantum information. Double quantum dots are the basic unit of such devices, since exchange between spins in neighboring dots provides two-qubit gates and Pauli spin blockade provides spin readout. Complex oxides offer a rich set of building blocks, including superconductivity, strong spin-orbit coupling and a high dielectric constant. Here we demonstrate single-electron devices at the \CZOKTO(100) interface, written by conductive atomic force microscope lithography. A single quantum dot shows well-resolved Coulomb diamonds, and sketched side gates act locally on the dot. In a double quantum dot, a combination of side-gate and back-gate tuning reveals a crossover from single-dot to double-dot behavior. Gating is able to tune through electron occupancies ($N_L$,$N_R$) that include (1,1), (2,0), and (0,2) in a second double quantum dot device which is relevant for singlet-triplet qubits. These results herald, in a complex oxide, the same transition from mesoscopic physics to quantum information.}


\maketitle

Spin qubits are generally made from group IV or III-V semiconductors ~\cite{10.1103/PhysRevA.57.120,10.1038/30156} whose heterostructure geometry enables top-down defining of quantum dots that can be exchange-coupled and sensed~\cite{10.1103/RevModPhys.79.1217,10.1103/RevModPhys.85.961,10.1103/RevModPhys.95.025003}. Each platform faces its own challenges, and some research directions involve exploration of hybrid materials that incorporate new properties, e.g., superconductivity, magnetism, or strong spin-orbit coupling via proximity effect~\cite{10.1038/nnano.2010.173,10.1038/s42254-020-0228-y}. Some semiconducting systems can be made superconducting themselves by doping, while others are intrinsically superconducting semiconductors~\cite{10.1103/PhysRevLett.12.474}.

The two-dimensional electron gases (2DEGs) formed at complex-oxide interfaces have been widely explored as a platform for quantum nanostructures using a variety of top-down nanolithographic techniques. Most of this work has been done at the LaAlO$_3$/SrTiO$_3$ (LAO/STO) interface. Conductive atomic force microscope (c-AFM) lithography can write and erase conducting nanostructures directly at the buried interface \cite{10.1038/nmat2136,10.1126/science.1168294}, and has been used to make single-electron transistors (SETs) \cite{10.1038/nnano.2011.56}, SETs in which electrons pair without forming a superconducting state \cite{10.1038/nature14398,10.1103/PhysRevX.6.041042}, and ballistic electron waveguides with quantized conductance \cite{10.1021/acs.nanolett.8b01614,10.1126/science.aat6467}. Nanoscale programming has also been achieved with ultra-low-voltage electron-beam lithography (ULV-EBL) \cite{10.1063/5.0027480}. Conventional electron-beam lithography has been used to define phase-coherent LAO/STO bridges \cite{10.1063/1.4768936}, and patterned growth and single-step nanolithography have produced one-dimensional quantum wires \cite{10.1103/PhysRevLett.112.136801} and islands that show resonant tunneling \cite{10.1103/PhysRevB.94.045120}. Local top and side gates have been used to create superconducting quantum interference devices \cite{10.1038/nnano.2016.112}, gate-tunable Josephson junctions \cite{10.1021/acs.nanolett.6b03820}, split-gate quantum point contacts \cite{10.1038/s41467-018-04657-z,10.1038/s41928-020-0383-2}, and quantum dots with pronounced Coulomb diamonds \cite{10.1038/s41467-017-00495-7,10.1103/PhysRevMaterials.4.122001}. Ballistic quantum point contacts with clean conductance quantization have also been made in SrTiO$_3$ using an ionic-liquid gate \cite{10.1038/s41928-023-00981-5}.

Despite this progress, oxide quantum dot devices have so far involved a single dot, leaving open whether two dots can be coupled and tuned independently, as spin qubits require. In sketched devices, gates behave non-locally, preventing the creation of double quantum dots and other coupled structures. Nethwewala \textit{et al.} found that gates 17 and 23~$\mu$m from a sketched waveguide tune it with only about ten times more voltage than side gates 2~$\mu$m away \cite{10.1002/admi.202400223}, a long-range, non-Coulombic coupling attributed to the ferroelastic domains of STO \cite{10.1038/nmat3810,10.1038/nmat3753} and also seen in frictional drag between nanowires \cite{10.1002/admi.201900301}. A gate that depletes an SET island therefore also depletes its leads and tunnel barriers, so the island cannot be emptied while transport through it is still measurable.

KTaO$_3$ (KTO) is, like STO, a quantum paraelectric with a very high dielectric permittivity ($\sim$5000 at low temperature) \cite{10.1143/JPSJ.41.888}, but it remains cubic, with no ferroelastic domains, and is much more uniform, local, and reproducible in its properties. KTO is also interesting because of its strong spin-orbit coupling, which arises from the Ta 5$d$ conduction band \cite{10.1002/aelm.201800860,10.48550/arXiv.2304.14426}, and because its interfaces host superconductivity. Superconductivity was found at (111) interfaces with $T_c$ up to about 2~K \cite{10.1126/science.aba5511,10.1126/science.abb3848}, and has since been observed at (110) \cite{10.1103/PhysRevLett.126.026802} and, most recently, at CaZrO$_3$/KTaO$_3$(001)~\cite{10.1021/acsnano.6c01438}, with $T_c$ up to about 0.25~K. These phenomena depend strongly on carrier density and crystallographic orientation \cite{10.1126/sciadv.adf1414,10.1063/5.0189956,10.1021/acs.nanolett.4c02500,10.1038/s41467-022-32242-y}. The low carrier density and high permittivity make KTO 2DEGs highly gate-tunable. Like STO, KTO can be programmed at the nanoscale using c-AFM~\cite{10.1021/acs.nanolett.2c00673,10.1103/PhysRevX.15.011037,10.1021/acs.nanolett.5c05590} and ULV-EBL~\cite{10.1021/acs.nanolett.2c00673}, and this approach has been used to write superconducting weak links, SQUIDs, and supercurrent diodes \cite{10.1103/PhysRevX.15.011037,10.1021/acs.nanolett.5c05590}. These results make KTO a promising platform for integrated superconducting and spin-orbit quantum devices \cite{10.48550/arXiv.2506.18894,10.1103/PhysRevX.15.011037,10.1021/acs.nanolett.5c05590}. Oxide quantum dots are attractive for spin-based quantum information, since the $d$-orbital character of the conduction band is expected to suppress the contact hyperfine interaction, and recent theory on LAO/STO double dots predicts high-fidelity spin SWAP operation at intermediate dot sizes \cite{10.48550/arXiv.2606.06948}. Gate-defined quantum dots have not yet been demonstrated in KTO, where the stronger spin-orbit coupling and the recently discovered superconductivity at the CZO/KTO(001) interface offer additional functionality.
Here, we demonstrate sketched single quantum dot and double quantum devices to single electron limit at the CZO/KTO(100) interface using c-AFM lithography technique. The double quantum dot is arguably more interesting as its occupation can be set one electron at a time starting from empty, and gate voltages can tune a double dot into a single dot. Such control of a double dot has not been achieved in sketched LAO/STO devices, despite extensive efforts, and is enabled here by the properties of KTO. Gate-defined quantum dots, including few-electron double and triple quantum dots, have recently been realized in ZnO heterostructures~\cite{10.1038/s41467-024-53890-2,10.1103/mnt5-s859,10.1038/s41598-025-20567-9}. Complex oxides such as KTO additionally offer a high dielectric constant, superconductivity, and strong spin-orbit coupling, which are new building blocks that can impact spin qubits.

\begin{figure}[!tb]
\centering
\includegraphics[width=\linewidth]{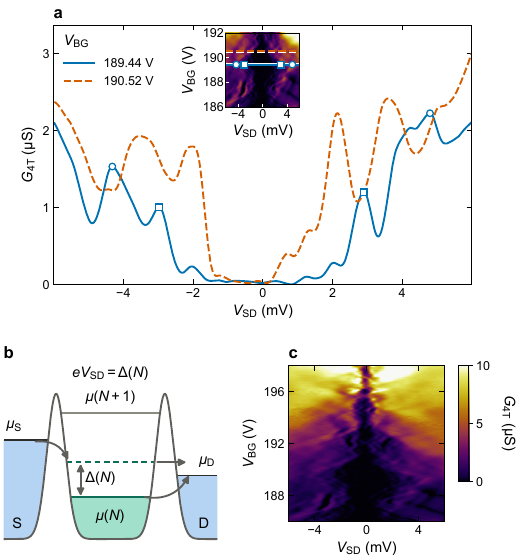}
\caption{Coulomb Blockade measurements from \DevA. (a) Differential conductance $G_{\mathrm{4T}}$ versus $V_{\mathrm{SD}}$ at $V_{\mathrm{BG}} = 189.44$~V (solid) and $190.52$~V (dashed), taken along the lines marked in the inset (a zoom of panel~(c)). Squares mark the onset of inelastic cotunneling at $|eV_{\mathrm{SD}}| = \Delta(N)$, and circles mark the onset of first-order tunneling at the diamond edges.
(b) Energy diagram at the onset of inelastic cotunneling, $eV_{\mathrm{SD}} = \Delta(N)$. The dot is in Coulomb blockade, with $\mu(N)$ below $\mu_{\mathrm{D}}$ and $\mu(N+1)$ above $\mu_{\mathrm{S}}$. An electron leaves the ground state for the drain while another enters the excited state from the source. Near the diamond edge, where $\mu(N)+\Delta(N)$ lies above $\mu_{\mathrm{D}}$, the electron in the excited state can then tunnel out to the drain by first-order tunneling (horizontal arrow), leaving $N-1$ electrons on the dot. Deeper in the diamond, $\mu(N)+\Delta(N)$ lies below $\mu_{\mathrm{D}}$ and this channel is closed.
(c) Differential conductance as a function of source-drain bias and back-gate voltage at zero magnetic field and a mixing-chamber temperature of 20~mK.}
\label{fig:SQD Diamonds}
\end{figure}

\begin{figure*}[tb!]
\centering
\includegraphics[width=\textwidth]{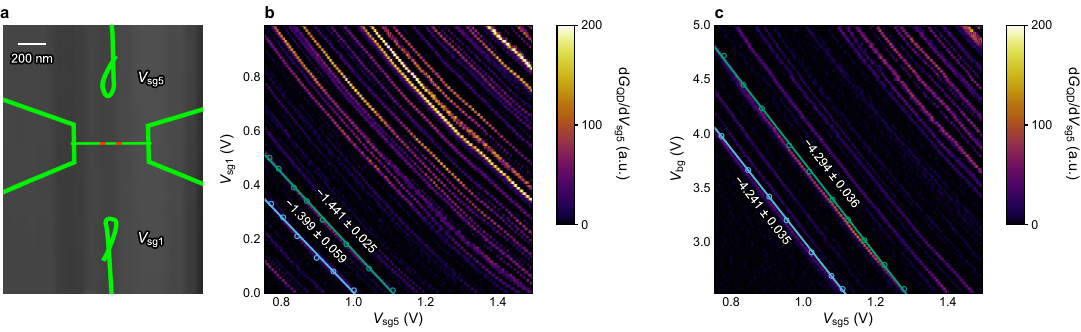}
\caption{\label{fig:FigureSQD_slopes}(a) Schematic showing the gates and device geometry of \DevB (a second single quantum dot). (b, c) Charge stability maps between pairs of gates, showing a lever-arm ratio of approximately $-1.4$ between $V_{sg1}$ and $V_{sg5}$ (b) and approximately $-4.2$ between $V_{bg}$ and $V_{sg5}$ (c).}
\end{figure*}

\begin{figure}[!tb]
\centering
\includegraphics[width=\linewidth]{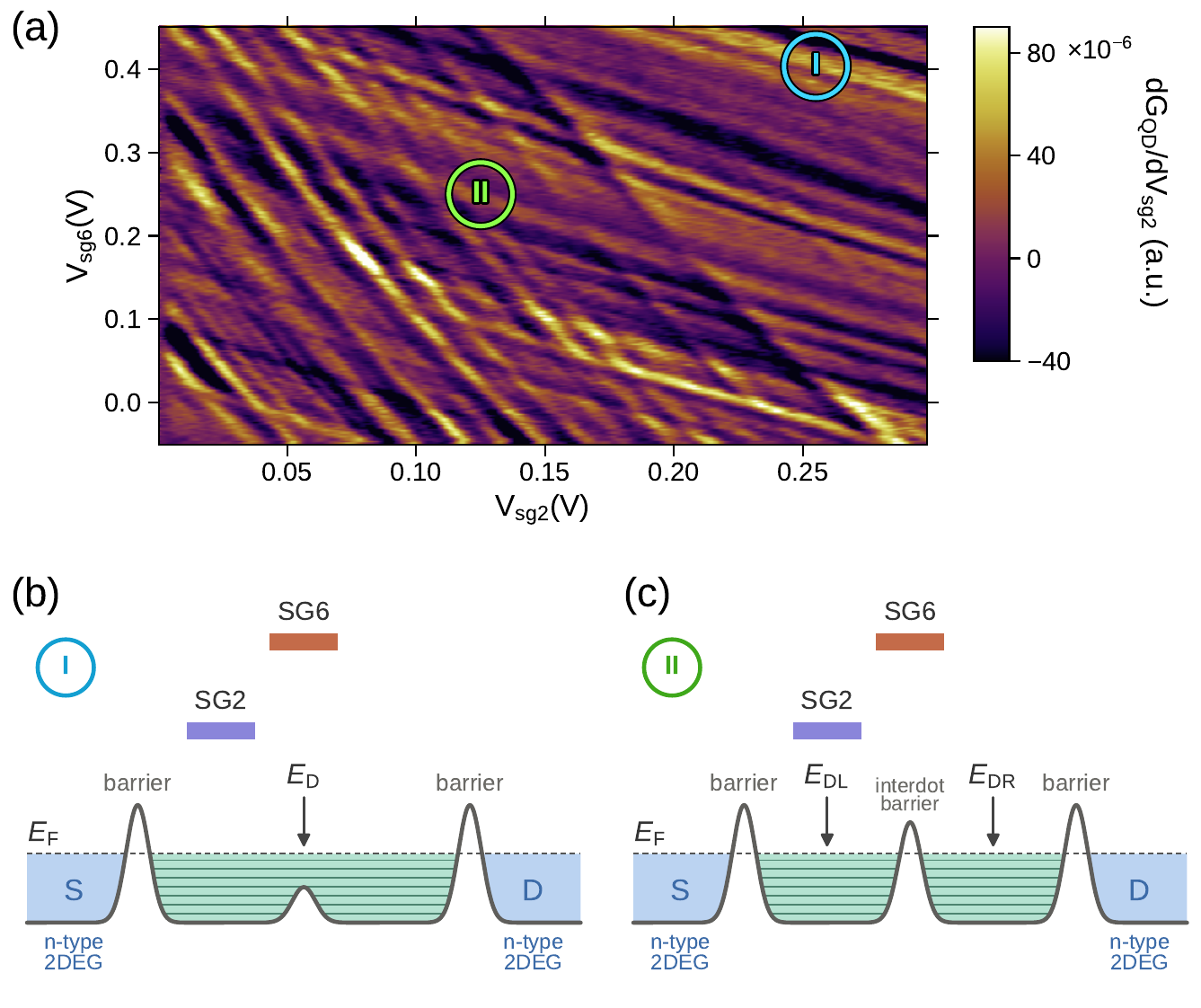}
\caption{Gate-tuned transition between a single dot and a double dot in \DevC. (a) Differential conductance $dG_\mathrm{QD}/dV_\mathrm{sg2}$ as a function of the side-gate voltages $V_\mathrm{sg2}$ and $V_\mathrm{sg6}$. At the higher gate voltages (region I), the charge transitions form a single family of parallel lines and at lower gate voltages (region II), two families with different slopes cross to form a Parallelogram-shaped cells. (b) Schematic conduction-band profile in region I: the gates GL and GR pull the band edge below the Fermi level $E_\mathrm{F}$ across the whole region between the source (S) and drain (D) barriers, forming one large dot with electrochemical potential $E_\mathrm{D}$. (c) Schematic profile in region II: a barrier between the gates rises above $E_\mathrm{F}$ and splits the well into two dots with electrochemical potentials $E_\mathrm{DL}$ and $E_\mathrm{DR}$.}
\label{fig:dqdB}
\label{fig:transport}
\end{figure}

\begin{figure*}[tb!]
\centering
\includegraphics[width=\textwidth]{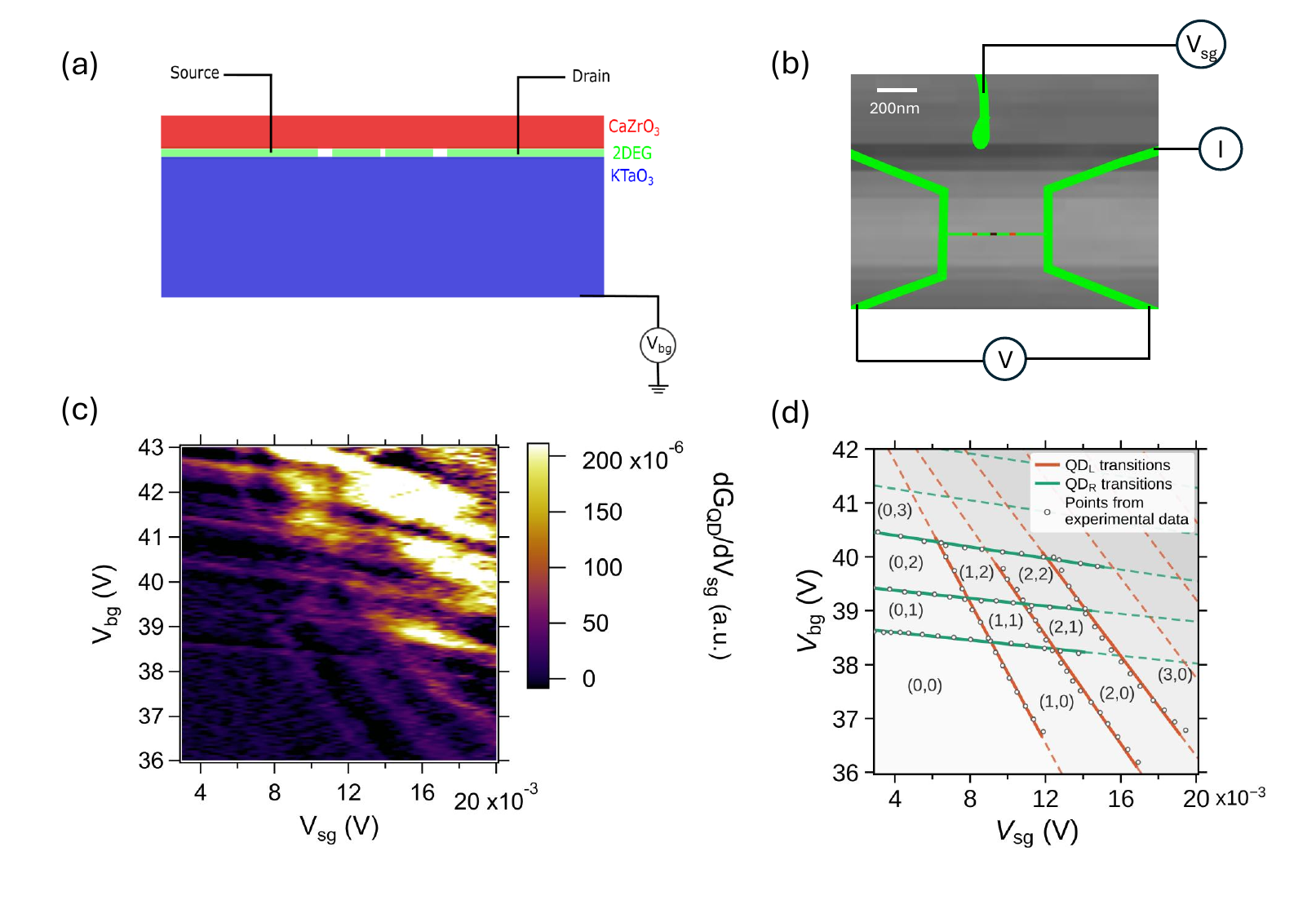}
\caption {Gate stability measurement from \DevD. (a),(b) Schematic of a written device: two islands, QD$_\mathrm{L}$ and QD$_\mathrm{R}$, between the source and drain leads, and a side gate (V$_\mathrm{sg}$) closer to QD$_\mathrm{L}$ and a global back gate is connected to the backside of KTO. (c) Charge stability diagram: $dG/dV_\mathrm{SG}$ as a function of the side-gate voltage $V_\mathrm{SG}$ and the back-gate voltage $V_\mathrm{BG}$. (d) Constant-interaction fit to the charge stability diagram in (c). Open circles are points from the experimental data; solid lines are straight-line fits to the QD$_\mathrm{L}$ (orange) and QD$_\mathrm{R}$ (green) transitions, and dashed lines continue the outermost fitted lines at their mean spacing. Labels give $(N_\mathrm{L}, N_\mathrm{R})$.
}
\label{fig:Device-D_charge_stability}
\end{figure*}

\section{Results}\label{sec:results}
\subsection{Device platform}\label{sec:platform}
A 6-nm-thick CZO film was grown on a KTO(100) substrate by pulsed laser deposition, using the approach developed for LaAlO$_3$/KTaO$_3$ heterostructures~\cite{10.1021/acs.nanolett.2c00673,10.1126/sciadv.adk4288}. CZO/KTO(001) heterointerfaces have been reported previously~\cite{10.1021/acsnano.6c01438}. The CZO was then removed by reactive ion etching everywhere except within a protected ``canvas'' region (Fig.~\ref{FigS2_define_canvas}). Within the canvas, conducting islands and tunnel barriers were defined by conductive atomic force microscope (c-AFM) lithography. We discuss two single-quantum-dot devices (\DevA, \DevB) and two double-quantum-dot devices (\DevC, \DevD) in the following sections.

\subsection{Coulomb blockade in a single quantum dot}\label{sec:sqd}
Figure~\ref{fig:SQD Diamonds} shows Coulomb blockade spectroscopy of \DevA, measured at a mixing-chamber temperature of 20~mK and zero magnetic field. At the charge degeneracy points, where neighboring diamonds would normally touch at $V_\mathrm{sd} = 0$, the conductance remains suppressed in a finite window around zero bias. This zero-bias gap shrinks with increasing back-gate voltage. The diamonds at higher back-gate voltage close nearly completely, approaching the conventional pattern, whereas those at lower back-gate voltage retain a finite gap.

The stability diagram also contains many lines running parallel to the diamond edges, both outside and inside the diamonds. Outside the diamonds, we attribute these lines to first-order tunneling through excited states of the dot, which sets in at a larger bias than first-order tunneling through the ground state at the diamond edges. Inside the diamonds, first-order tunneling is blocked and transport proceeds by cotunneling~\cite{10.1103/PhysRevLett.86.878}. Elastic cotunneling returns the dot to its ground state and produces a weak, nearly featureless background at low bias. Inelastic cotunneling leaves the dot in an excited state at energy $\Delta(N)$ above the ground state. It therefore sets in at $|eV_\mathrm{sd}| = \Delta(N)$, independent of gate voltage. The rates of both processes scale with the product of the source and drain tunnel couplings, so cotunneling features become more pronounced as the dot--lead coupling increases. Cotunneling alone does not produce lines parallel to the diamond edges inside the diamonds. These lines are instead consistent with first-order tunneling through an excited state that has been populated by inelastic cotunneling~\cite{10.1103/PhysRevLett.94.206805}.

\subsection{Local gating with sketched side gates}\label{sec:localgating}
We performed gate stability measurements between different gates of a second single-quantum-dot device (\DevB). For \DevB, SG1 was written approximately 1.6~$\mu$m and SG5 approximately 1.2~$\mu$m from the quantum dot.
In the charge stability diagram as a function of the two gate voltages, all transitions form a single family of parallel lines. The lever-arm ratio between $V_{sg1}$ and $V_{sg5}$ is approximately $-1.4$, which shows the capability of this platform for local gating. The lever-arm ratio between $V_{bg}$ and $V_{sg5}$ is approximately $-4.2$. The sketched in-plane gates act as local gates. Their coupling strength decreases as the gates are placed farther from the quantum dot (Fig.~\ref{fig:FigureSQD_slopes}), and this trend holds across different cooldowns.

\subsection{Gate-tuned transition between single and double dots}\label{sec:dqdC}
We next discuss two double-quantum-dot devices (\DevC, \DevD). In \DevC, the two side gates can fill the dots enough to merge them into a single dot. The $\mathrm{d}G_{\mathrm{QD}}/\mathrm{d}V_{\mathrm{sg2}}$ map in Fig.~\ref{fig:dqdB}(a) was measured at a base temperature of 50~mK in the absence of a magnetic field and contains two sets of charge-transition lines with clearly different slopes. The electrochemical potential of a dot is constant along its transition lines, so the slope $\mathrm{d}V_{\mathrm{sg6}}/\mathrm{d}V_{\mathrm{sg2}} = -\alpha_{\mathrm{sg2}}/\alpha_{\mathrm{sg6}}$ is the ratio of that dot's lever arms to the two side gates. Straight-line fits give median slopes of $-3.25$ for the steeper set and $-0.99$ for the shallower set.
SG2 was sketched closer to the left dot (approximately 200~nm away), while SG6 is farther away and centered on the interdot barrier (approximately 900~nm away). The left dot couples to SG2 about three times more strongly than to SG6. The shallower lines, with a lever-arm ratio close to one, are assigned to the right dot. SG2 is laterally offset from the right dot, and SG6, although laterally closer, sits farther from the channel, so the two gates couple to the right dot with comparable strength. Raising $V_{\mathrm{sg2}}$ and $V_{\mathrm{sg6}}$ fills both dots and lowers the interdot barrier, which is closest to SG6. The device therefore evolves from two crossing sets of lines in regime II to a single set of parallel lines in regime I, where the barrier lies below $E_{\mathrm{F}}$ and the two dots behave as one (Fig.~\ref{fig:dqdB}(b,c)).

\subsection{Sequential loading of a double quantum dot from (0,0) state}\label{sec:dqdD}
\DevD was sketched to have approximately two conducting islands with 100 nm barriers at room temperature. In this device, the side gate is positioned closer to QD$_\mathrm{L}$, while the back gate acts on both dots. Because the two gates couple to the dots in different ratios, together they control the occupation of each dot independently. We explore the low-occupation limit at 50~mK in an out-of-plane magnetic field of 12~T (Fig.~\ref{fig:Device-D_charge_stability}). The in-plane side gate acts strongly on the adjacent QD$_\mathrm{L}$, whereas the back gate acts weakly on both dots and couples more strongly to QD$_\mathrm{R}$. This asymmetry may reflect a difference in the conducting areas of the two islands, or partial screening of QD$_\mathrm{L}$ from the back-gate field by the nearby side gate. The two dots therefore produce two families of transitions (Fig.~\ref{fig:Device-D_charge_stability}), whose slopes are set by how strongly each gate acts on each dot.

The two dots produce two families of transitions: nearly horizontal lines with $dV_\mathrm{BG}/dV_\mathrm{SG} \approx -0.036$~V/mV, which we assign to the larger QD$_\mathrm{R}$, and steep lines with slopes between $-0.50$ and $-0.65$~V/mV, which we assign to the smaller QD$_\mathrm{L}$. The steep lines reflect the strong coupling of QD$_\mathrm{L}$ to the adjacent side gate.
No transitions appear below the lowest horizontal line or to the left of the leftmost vertical line, so we assign the lower-left region to $(N_\mathrm{L},N_\mathrm{R})=(0,0)$; the two dots are therefore filled with one electron at a time starting from empty. Together, the two double-dot devices show that gate voltages alone set both the occupation and the configuration of the dots: in one device, a side gate and the back gate load each dot one electron at a time starting from empty, and in the other, two side gates tune the system from the double-dot to the single-dot regime by lowering the barrier between the dots.

\section{Discussion and Outlook}\label{sec:discussion}
Coulomb blockade spectroscopy of \DevA shows gaps at the charge degeneracy points. Although similar gaps have been attributed to negative-$U$ pairing in LAO/STO~\cite{10.1038/nature14398,10.1038/s41467-017-00495-7}, we do not observe a characteristic diamond-doubling transition in the KTO(100) devices (Fig.~\ref{fig:allfields}). Either local pairing is absent at this interface or its energy scale lies below our resolution; in both cases, preformed pairs may not be a generic feature of quantum-paraelectric oxide interfaces. The zero-bias gap in the single-dot devices can instead arise from a small accidental dot in series with the main dot. Its large addition energy and weak coupling to the gate keep it Coulomb-blockaded across many of the main dot's diamonds, so current flows only once the bias pulls one of its levels into the bias window. The gap closes at higher gate voltage as the barriers of the accidental dot become transparent and it merges with the lead, in the same way that the intentional double dots become single dots (Fig.~\ref{fig:dqdB}, region~I).\\
The two double-dot devices probe different regimes of interdot coupling. In \DevC, the side gates tune the interdot barrier, so the same device operates as either a single dot or a double dot. \DevD remains in the weakly coupled limit, where the two dots behave almost independently. The weak coupling also explains why the stability diagram of \DevD consists of parallelogram-shaped cells rather than a resolved honeycomb. The short interdot segments of a honeycomb, whose length is set by the interdot capacitive coupling, are too short to resolve here, and the cells are sheared because both gates act on both dots. The weak coupling may reflect the separation of the written islands or screening by the surrounding conductors and the high-permittivity substrate. 

The two-electron (2,0) and (0,2) states resolved in \DevD (Fig.~\ref{fig:Device-D_charge_stability}) are directly relevant to singlet-triplet qubits~\cite{10.1103/PhysRevLett.89.147902,10.1126/science.1116955}. When both electrons occupy the same dot, the exchange coupling is large, whereas in the (1,1) configuration it is small, so pulsing the detuning between these charge states switches the exchange on and off. This mode of operation has been demonstrated in GaAs~\cite{10.1126/science.1116955,10.1038/nphys1424} and Si/SiGe~\cite{10.1038/nature10707} double dots, and the exchange can also be controlled with a barrier gate between the dots~\cite{10.1103/PhysRevLett.116.116801}. 
The high dielectric permittivity of KTO is also expected to screen charge noise, which limits exchange-based qubits through fluctuations of the detuning~\cite{10.1103/PhysRevLett.116.116801}. A charge fluctuator near the dot is screened by the substrate, and the low microwave loss of KTO~\cite{10.1063/1.1905789} implies little intrinsic dielectric noise. The interfacial electric field reduces the permittivity near the dot, and charge noise in written KTO devices remains to be measured. Several steps remain before the KTO-based double dots can be operated as qubits. 

Our observations extend the c-AFM device toolbox to KTO and are a first step toward spin and hybrid superconductor-quantum dot devices in a strong spin-orbit oxide.

\section{Methods}\label{sec:methods}

\subsection*{Sample and device fabrication}
The c-AFM technique provides great versatility in the creation of nanoscale devices~\cite{10.1038/nnano.2011.56,10.1021/acs.nanolett.2c00673,10.1021/acs.nanolett.5c05590} on both STO and KTO platforms. The key device element is a nanoscale tunnel barrier or weak link that is integral to many devices, for example, sketched SETs~\cite{10.1038/nnano.2011.56} and sketched SQUIDs~\cite{10.1103/PhysRevX.15.011037}. The CZO/KTO interface was contacted using aluminum wire bonds. The interface is inherently conducting at room temperature. In this work, 60--80~nm conducting islands were created by c-AFM lithography using an Asylum Research MFP-3D AFM. The AFM chamber was maintained at an ambient temperature of 25\textdegree C and a humidity of about 40\%.

The c-AFM technique is used to make part of the CZO/KTO interface insulating and then revert only the desired areas to be conducting (Fig.~\ref{FigS2_define_canvas}). This is done by applying a negative or positive dc voltage to the AFM tip ($V_{tip}$) through a 1~G$\Omega$ tip resistor. To erase the core region of the canvas ($\sim 60~\mu\mathrm{m} \times 60~\mu\mathrm{m}$), $V_{tip}$ values from $-7$ to $-14$~V were used, and to write the virtual leads, $V_{tip}$ was in the range $+26$ to $+32$~V. The main channels of the quantum dot devices were written with $V_{tip}$ of $+7$ to $+8$~V, and two tunnel barriers were defined by cutting the main channel perpendicularly with $V_{tip}$ of $-2$ to $-2.5$~V. For the double quantum dot device, three tunnel barriers were created using a similar voltage range. The conductance was monitored throughout the entire writing process.

\backmatter

\bmhead{Supplementary information}
Supplementary Figs.~S1--S3 and Supplementary Table~S1 are provided in the Supplementary Information section below.

\bmhead{Acknowledgements}
This material is based upon work supported by the U.S. Department of Energy, Office of Science, Office of Basic Energy Sciences, under Award Number DE-SC0026132.

\bmhead{Author contributions}
S.B. processed the samples, fabricated the devices, performed the measurements, analyzed the data, and wrote the manuscript. J.Y. and P.M. grew the CZO/KTO heterostructures. M.K. contributed to the data analysis. S.T. contributed to device fabrication and to editing the manuscript. R.R. contributed to sample processing and measurements. A.H contributed to measurements. P.I. developed the measurement software. C.-B.E. supervised the sample growth. J.L. supervised the project and contributed to writing the manuscript.

\bmhead{Notes}
During the submission of this manuscript, a related complex oxide double quantum dot work was reported by Meucci et al\cite{Meucci2026-op}.
The authors declare no competing interests.

\bmhead{Data availability}
The data shown in this paper will be provided with this paper. Additional data that support the plots and other analysis in this work are available from the
corresponding author upon request. 

\bibliography{references,paperpile}

\clearpage
\setcounter{figure}{0}
\setcounter{table}{0}
\renewcommand{\thefigure}{S\arabic{figure}}
\renewcommand{\thetable}{S\arabic{table}}

\section*{Supplementary Information}\label{sec:SI}

\begin{figure}[!htb]
\centering
\includegraphics[width=\textwidth]{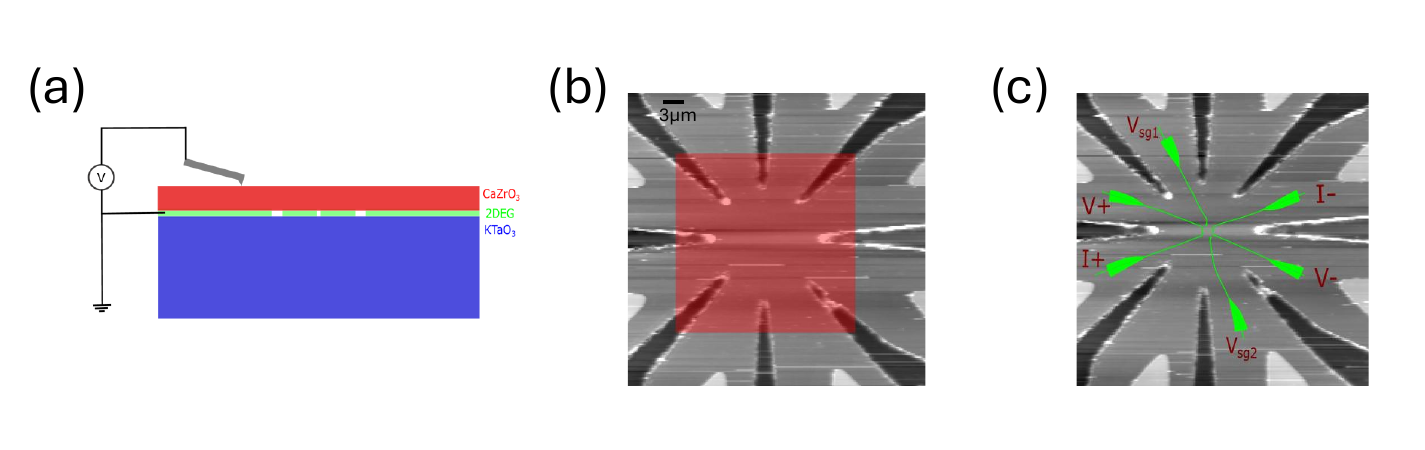}
\caption{ (a) Schematic cross-section of the device and of c-AFM lithography. A biased AFM tip scanned over the CaZrO$_3$ surface writes conducting regions of the 2DEG at the CZO/KTO interface, and gaps in the written pattern form the tunnel barriers. (b) AFM height image of a typical CZO/KTO ``canvas''. Dark regions are where the CZO has been etched, and the light grey region is where the conducting 2DEG is present. The red-shaded region is made insulating by scanning it with a negatively biased AFM tip, a process referred to as erasing. (c) The green structures are virtual leads, written with a positively biased tip, which connect the device to the conducting 2DEG outside the erased area.}
\label{FigS2_define_canvas}
\end{figure}

\begin{figure}[!htb]
\centering
\includegraphics[width=\textwidth]{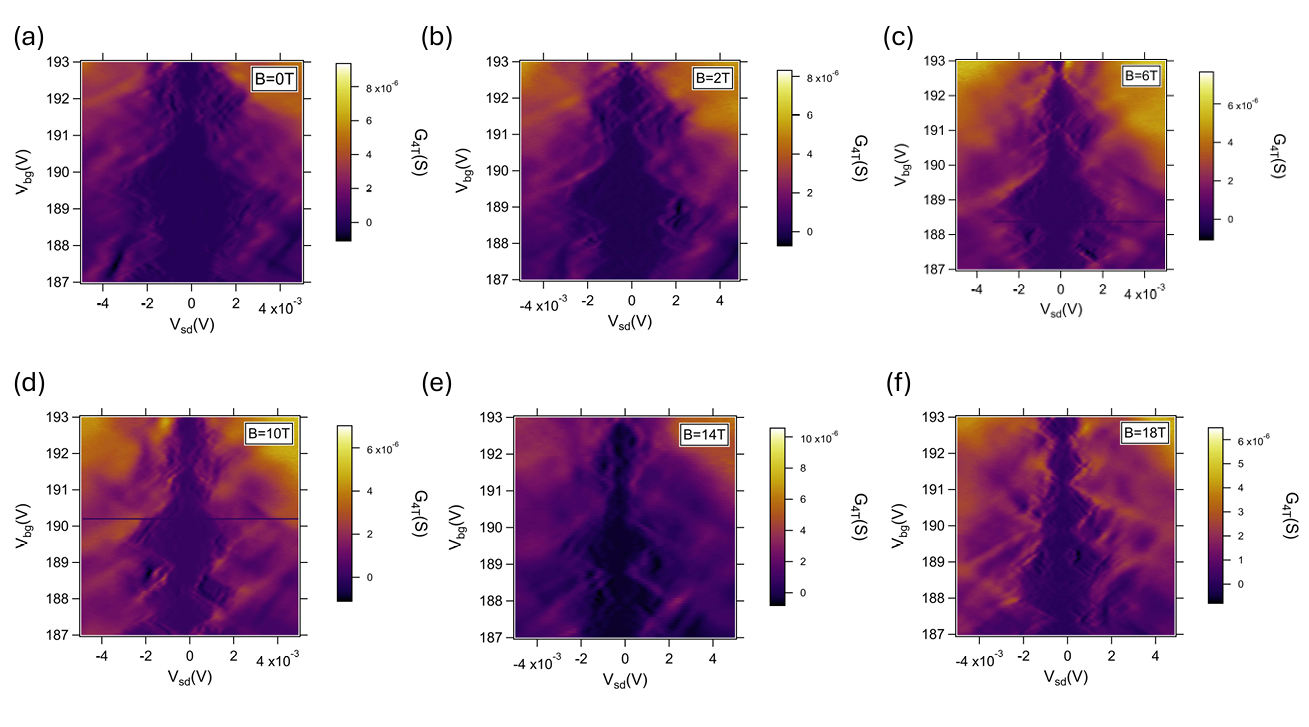}
\caption{\label{fig:allfields} Conductance map of device S1 as a function of source-drain bias and back-gate voltage at magnetic fields from 0 to 18~T and a temperature of 20~mK.}
\end{figure}

\begin{table}[!htb]
\centering
\caption{Coulomb-diamond parameters extracted from the 18\,T stability diagram.}
\label{tab:diamond_params}
\begin{tabular}{lcccccc}
\toprule
Diamond & $E_C$ (meV) & $C_\Sigma$ (aF) & $C_s$ (aF) & $C_d$ (aF) & $C_g$ (aF) & $\alpha$ \\
\midrule
D1 & 0.997 & 160.70 & 75.90 & 85.80 & 0.1214 & $7.55\times10^{-4}$ \\
D2 & 0.646 & 248.21 & 120.63 & 127.57 & 0.1381 & $5.56\times10^{-4}$ \\
D3 & 0.719 & 222.83 & 95.60 & 127.24 & 0.1821 & $8.17\times10^{-4}$ \\
D4 & 0.451 & 355.25 & 261.81 & 93.44 & 0.1907 & $5.37\times10^{-4}$ \\
\botrule
\end{tabular}
\end{table}

\end{document}